\documentclass[twocolumn,showpacs,preprintnumbers,amsmath,amssymb,superscriptaddress]{revtex4}

\usepackage{graphicx}
\usepackage{dcolumn}
\usepackage{bm}
\newcommand{\qubit}[1]{| #1 \rangle}

\newcommand{\inner}[2]{\langle #1 | #2 \rangle}
\newcommand{\density}[2]{| #1 \rangle \langle #2 |}

\newcommand{\partdif}[1]{\frac{\partial}{\partial #1}}

\usepackage{amsmath, amssymb, theorem}
\theoremstyle{break}
\theorembodyfont{\normalfont}
\newtheorem{theorem}{Theorem}

\newtheorem{lemma}{Lemma}

\newtheorem{remark}{Remark}

\newenvironment{proof}{\noindent Proof.\  \\}{\hspace*{\fill}$\square$ \vspace{2ex}}

\begin{document}
\preprint{APS/123-QED}

\title{
 A Quantum Key Distribution Protocol with Selecting Announced States, \\ 
 Robust against Photon Number Splitting Attacks
}

\author{EGUCHI Makoto}
 \affiliation{
  SHARP Corporation,
  22-22 Nagaike-cho, Abano-ku, Osaka-shi, Osaka, Japan
 }
\author{HAGIWARA Manabu}
 \affiliation{
  National Institute of Advanced Industrial Science and Technology,
  1-18-13 Sotokanda, Chiyoda-ku, Tokyo, Japan
 }
\author{Hideki IMAI}
 \affiliation{
  National Institute of Advanced Industrial Science and Technology,
  1-18-13 Sotokanda, Chiyoda-ku, Tokyo, Japan
 }
 \affiliation{
  Institute of Industrial Science, University of Tokyo,
  4-6-1 Komaba, Meguro-ku, Tokyo, Japan
 }
\date{\today}

\begin{abstract}
We propose a new class of quantum key distribution protocol,
that ended up to be robust against photon number splitting attacks
in the weak laser pulse implementations.
This protocol comprises of BB84 protocol and SARG protocol,
especially in aspects of controlling classical sifting procedures of two protocols.
The protocol is more secure than both of BB84 protocol and SARG protocol,
and the ultimate limit of robustness in the proposed protocol expands as well than both of them.
\end{abstract}

\pacs{Valid PACS appear here}

\maketitle

\section{Introduction}

Quantum Key Distribution (QKD) protocol is the only physically secure method
for the distribution of a secret key between two distant partners (called \textit{Alice} and \textit{Bob}).
The physical secure comes from the well-known facts
that an attacker (called \textit{Eve}) cannot measure
an unknown quantum state without modifies the state itself,
and she cannot duplicate the state and forward a perfect copy to Bob.
The facts are proved by two principles, ``Uncertainty principle'' and ``No cloning theorem''.
BB84 protocol \cite{BB84} is the first single-photon QKD protocols,
which use a random string of signal states which, for example, can be realized as single photons
in horizontal, vertical, right circular or left circular polarization states.

In recent years, several long-distance implementations of BB84 protocol have been developed,
that use photons as information carriers and optical fibers as quantum channels.
Most often Alice sends to Bob a coherent weak laser pulse in which she has encoded the bit.
In weak pulses QKD system, there are the pulses which contains more than one photon
with non-negligible probability.
It implies that for these pulses Eve no longer limited by ``No cloning theorem'',
and she can perform new types of attacks to obtain the secret key without introducing errors.
In such the attacks, there are Photon Number Splitting (PNS) attacks \cite{AGS04}\cite{Lu00}\cite{SARG04}.
Although PNS attacks are far beyond today's technology,
if one includes them in the security analysis, the consequences are dramatic
and long-distance weak laser pulse QKD systems no longer have physical security.

In this paper, we propose a new QKD protocol robust against PNS attacks,
achieved by alternative of BB84 protocol and SARG protocol \cite{SARG04}.
The protocol is more secure than both of BB84 protocol and SARG protocol,
especially long-distance weak laser pulses QKD systems.
The advantage of this protocol is that it is easy to implement,
just because it is composed of an existing quantum key distribution system,
where the classical sifting procedure is only changed
which is easier than making a perfect single-photon source.

\section{Proposed Protocol}

Our protocol uses four quantum states
\begin{equation*}
 \mathcal{Q} := \{ \qubit{+x}, \qubit{-x}, \qubit{+z}, \qubit{-z} \}
\end{equation*}
such that $| \inner{\omega x}{\omega' z} | = 1 / \sqrt{2}$ with $\omega, \omega' \in \{ +, - \}$
and $| \inner{+ \alpha}{- \alpha} | = 0$ with $\alpha \in \{ x, z \}$.
The four states are also used by BB84 protocol and SARG protocol.
$\qubit{\pm x}$ and $\qubit{\pm z}$ denote
the eigenvectors of $\sigma_x$ and $\sigma_z$ with eigenvalue $\pm 1$, respectively.

Our protocol contains following phases;

\begin{enumerate}
\item[1]
 Quantum communication phase \\
 Alice selects randomly one of four states $\qubit{A} \in \mathcal{Q}$ and 
 sends $\qubit{A}$ to Bob.
 Bob measures either $\sigma_x$ or $\sigma_z$,
 and gets a state $\qubit{B} \in \mathcal{Q}$.
 We call $\qubit{A}$ and $\qubit{B}$ \textit{raw keys}.
\item[2]
 Selecting annoucement phase \\
 Alice performs a procedure, in which she obtains $0$ with the probability $a$,
 and $1$ with the probability $1 - a$, and she gets $A \in \{ 0, 1 \}$.
 The probability $a$ is determined uniquely by the length of fiber and $0 \le a \le 1$.
 If $A = 0$, go to step 3-1 and 4-1, and otherwise, go to step 3-2 and 4-2.
\item[3-1]
 Classical announcement phase (for $A = 0$) \\
 Alice announces publicly a pair of two states
 $\mathcal{A} = \{ \qubit{A_1}, \qubit{A_2} \}$, 
 such that $\qubit{A} \in \mathcal{A}$ and $|\inner{A_1}{A_2}| = 0$.
 It means that Alice announces a pair of orthogonal states.
\item[4-1]
 Sifting and decoding phase (for $A = 0$) \\
 When $\qubit{B} \in \mathcal{A}$,
 they get bits, called \textit{sifted keys}, from $\qubit{A}$ and $\qubit{B}$ with the convention that
 $\qubit{+x}$ and $\qubit{+z}$ code for $0$ and $\qubit{-x}$ and $\qubit{-z}$ code for $1$. \\
 When $\qubit{B} \notin \mathcal{A}$, they discard their raw keys.
\item[3-2]
 Classical announcement phase (for $A = 1$) \\
 Alice selects randomly one of two pairs of states
 $\mathcal{A} = \{ \qubit{A_1}, \qubit{A_2} \}$,
 such that $\qubit{A} \in \mathcal{A}$ and $|\inner{A_1}{A_2}| = 1 / \sqrt{2}$,
 and announces publicly $\mathcal{A}$ to Bob.
 It means that Alice announces a pair of nonorthogonal states.
\item[4-2]
 Sifting and decoding phase (for $A = 1$) \\
 When $\qubit{B} \notin \mathcal{A}$,
 Bob obtains $\qubit{B'}$ from $\qubit{B}$,
 such that $\qubit{B'} \in \mathcal{A}$ and $| \inner{B}{B'} | = 1 / \sqrt{2}$,
 and they get sifted keys from $\qubit{A}$ and $\qubit{B'}$ with the convention that
 $\qubit{\pm x}$ code for $0$ and $\qubit{\pm z}$ code for $1$. \\
 When $\qubit{B} \in \mathcal{A}$, they discard their raw keys.
\end{enumerate}

\begin{remark}
BB84 is described as the proposed protocol with $a = 1$,
and SARG is same as this protocol with $a = 0$.
\end{remark}

\section{Photon Number Splitting Attacks}

In weak pulses QKD system, Alice sends to Bob a weak laser pulse in which she has encoded the bit.
Each pulse is a priori in a coherent state of weak intensity,
which can be rewritten as a mixture of Fock states, $\sum_{n \ge 0} p_n \density{n}{n}$,
with the number $n$ of photons distributed according to the Poissonian statistics of mean $\mu$,
$p_n = e^{\mu} / n!$ \cite{AGS04}\cite{Lu00}\cite{SARG04}.

Consider now the implementation of the proposed protocol with weak pulses.
Bob's detector is triggered with probability,
taking into account intensities of weak laser pulses,
channel losses and imperfect detection efficiencies.
Then, in the absence of Eve, Bob's raw detection rate,
which is the probability that he detects a photon per pulse sent by Alice, is given by
\begin{equation*}
 R_{\mathrm{raw}}(\eta_{\rho})
  = \sum_{n \ge 1} p_n \{ 1 - (1 - \eta_{\mathrm{d}} \eta_{\rho})^n \}
  \simeq \eta_{\mathrm{d}} \eta_{\rho} \mu
\end{equation*}
where $\eta_{\mathrm{d}}$ is the quantum efficiency of a detector and
$\eta_{\rho}$ is the channel transmission.

In this case, if we endow Eve with unlimited technological power within the laws of quantum physics,
the following attacks, named a \textit{storage attack} and
an \textit{Intercept Resend with Unambiguous Discrimination attack} (shortly an \textit{IRUD attack}),
are possible in principle \cite{SARG04}.
(We will explain details of these attacks later.)  
If Alice and Bob are connected by a lossy channel ($\eta_{\rho} < 1$)
and Eve has a lossless channel ($\eta_{\rho} = 1$) which connects Alice and Bob,
Eve performs either attacks on a fraction $q$ of pulses, that is, she tries as follows:
\begin{enumerate}
\item
 Eve performs a procedure, in which she obtains $0$ with the probability $q$
 and $1$ with the probability $1 - q$.
\item
 When she gets $1$, she only forwards the pulse to Bob using her lossless channel.
 When she gets $0$, she performs one of the two PNS attacks.
\end{enumerate}
The attack probability $q$ depends on both a type of her attack and the length of lossy channel,
such that Alice and Bob do not notice any change in the expected raw rate and Eve remains undetected.

\subsection{Storage Attack}

We will explain the procedure of a storage attack \cite{Lu00} in the following.
\begin{enumerate}
\item
 Eve counts the number of photons in the pulse, using photon number quantum nondemolition measurement.
 If the pulse contains only one photon, Eve discards the photon.
\item
 When Eve detects that it is a multiphoton pulse,
 she keeps one of the photons in a quantum memory and  forwards the remaining photons to Bob,
 using a perfectly transparent quantum channel, $\eta_{\rho} = 1$.
\item
 By the information in classical announcement phase,
 Eve correspondingly measures the photon stored in her quantum memory.
\end{enumerate}

When Eve applies a storage attack on a fraction $q$ of the pulses,
Bob's raw detection rate is
\begin{equation*}
\begin{split}
 R^{S}(q)
  &= (1 - q) \eta_{\mathrm{d}} \mu
   + q \sum_{n \ge 2} p_n \{ 1 - (1 - \eta_{\mathrm{d}})^{n - 1} \} \\
  &\simeq (1 - q) \eta_{\mathrm{d}} \mu + q \eta_{\mathrm{d}} p_2 \text{.}
\end{split}
\end{equation*}
By Lemma \ref{lemma:info}, her mutual information of the key is
\begin{equation*}
 I^{S}_{\mathrm{Pr}}(q)
  = \frac{p \eta_{\mathrm{d}} p_2}{(1 - q) \eta_{\mathrm{d}} \mu + q \eta_{\mathrm{d}} p_2}
  \cdot I^{S}_{a}
\end{equation*}
where
\begin{equation*}
 I^{S}_{a} = 1 - (1 - a) \cdot H \left(\frac{\sqrt{2} + 1}{2 \sqrt{2}} \right)
\end{equation*}
with $H(x) = - x \log_2 x + (1 - x) \log_2 (1 - x)$.

\begin{lemma}[\cite{Per98}]
 Eve is now faced with the problem of detecting two states ($\qubit{x}$ and $\qubit{y}$), 
 having an overlap $| \inner{x}{y} | = \chi$.
 Then she applies the measurement maximizing her information, obtaining
\begin{equation*}
 I(\chi) = 1 - H(P)
\end{equation*}
where $P = \frac{1}{2} (1 + \sqrt{1 - \chi^{2}})$.
\label{lemma:info}
\end{lemma}

Given $\eta_{\rho}$, Eve chooses $q$ such that $R_{\mathrm{raw}}(\eta_{\rho}) = R^{S}(q)$
and her mutual information of the sifted key is
\begin{equation*}
 I^{S}_{\mathrm{Tr}}(\eta_{\rho})
  = ({\eta_{\rho}}^{-1} - 1) \cdot \frac{\mu}{{p_2}^{-1} - 1} \cdot I^{S}_{a} \text{.}
\end{equation*}

\subsection{Intercept Resend with Unambiguous Discrimination Attack}
An encoded pulse containing three photons is rewritten as one of the four states
\begin{equation*}
\begin{split}
 &\left\{ \qubit{\Psi_1}, \qubit{\Psi_2}, \qubit{\Psi_3}, \qubit{\Psi_4} \right\} \\
 & \qquad = \left\{ \qubit{+x}^{\otimes 3}, \qubit{-x}^{\otimes 3},
                    \qubit{+z}^{\otimes 3}, \qubit{-z}^{\otimes 3} \right\} \text{.}
\end{split}
\end{equation*}
In this case, there exist four orthogonal states of three qubits,
$\qubit{\Phi_1}, \ldots, \qubit{\Phi_4}$,
such that $\inner{\Psi_i}{\Phi_j} = \delta_{ij} \frac{1}{\sqrt{2}}$.
Therefore, we can perform a measurement $\mathcal{M}$, that distinguishes unambiguously
among $\qubit{\Psi_1}, \ldots, \qubit{\Psi_4}$, with a probability of success $p_{\mathrm{ok}} = 1 / 2$.

We will explain the procedure of an IRUD attack \cite{SARG04} in the following.
\begin{enumerate}
\item
 Eve measures the number of photons and discards a pulse containing less than three photons.
\item
 On the pulse containing at least three photons, Eve performs the measurement $\mathcal{M}$.
\item
 If the result is conclusive, she sends a new photon prepared in the good state to Bob
 using a perfectly transparent quantum channel.
 If not conclusive, Eve discards the result and the pulse.
\end{enumerate}

When Eve applies the IRUD attack on a fraction $p$ of the pulses,
Bob's raw detection rate and Eve's mutual information are
\begin{equation*}
\begin{split}
 R^{I}(q)
  &= (1 - q) \eta_{\mathrm{d}} \mu
   + q p_{\mathrm{ok}} \sum_{n \ge 3} p_n \{ 1 - (1 - \eta_{\mathrm{d}})^{n - 2} \} \\
  &\simeq (1 - q) \eta_{\mathrm{d}} \mu + q \eta_{\mathrm{d}} p_{\mathrm{ok}} p_3
\end{split}
\end{equation*}
and
\begin{equation*}
 I^{I}_{\mathrm{Pr}}(q)
  \simeq \frac{q \eta_{\mathrm{d}} p_{\mathrm{ok}} p_3}
               {(1 - q) \eta_{\mathrm{d}} \mu + q \eta_{\mathrm{d}} p_{\mathrm{ok}} p_3} \text{.}
\end{equation*}

When Eve chooses $q$ such that $R_{\mathrm{raw}}(\eta_{\rho}) = R^{I}(q)$,
her mutual information of the sifted key is
\begin{equation*}
  I^{I}_{\mathrm{Tr}}(\eta_{\rho})
   = ({\eta_{\rho}}^{-1} - 1) \cdot \frac{1}{{(p_{\mathrm{ok}} p_3)}^{-1} - 1} \text{.}
\end{equation*}

\section{Security Against PNS Attacks}

In this section, we evaluate security against PNS attacks with $\mathrm{QBER} = 0$.
In proposed protocol, the sifted key rate,
which is the probability that Alice and Bob share a sifted key per a pulse, is given by
\begin{equation*}
 R_{\mathrm{sift}}(a, \eta_{\rho})
  \simeq \frac{1 + a}{4} \cdot \eta_{\mathrm{d}} \eta_{\rho} \mu \text{.}
\end{equation*}

It is easy to see that security against PNS attacks will be decreasing the sifted key rate.
Therefore, we shall evaluate a security under the condition
that a sifted key rate is constant regardless of the selecting probability $a$ \cite{SARG04}.
Then, we change $\mu$ to
\begin{equation*}
 \mu_a = \frac{2}{1 + a} \cdot \mu_B
\end{equation*}
where $\mu_B$ is the mean photon number when using BB84 protocol.
In this paper, we use a typical value $\mu_B = 0.1$.

Eve's mutual information of the sifted key when she performs either of two PNS attacks is resprctively
\begin{equation*}
 I^{S}(a, \eta_{\rho})
  = ({\eta_{\rho}}^{-1} - 1) \cdot \frac{1}{\frac{2}{e^{- \mu_a} \mu_a} - 1} \cdot I^{S}_{a}
\end{equation*}
and
\begin{equation*}
 I^{I}(a, \eta_{\rho})
  = ({\eta_{\rho}}^{-1} - 1) \cdot \frac{1}{\frac{12}{e^{- \mu_a} {\mu_a}^2} - 1}
\end{equation*}

From these equations, we have the following theorem:

\begin{theorem}
 Consider Alice and Bob share a secret key using weal laser pulse QKD system and our proposed protocol.
 They choose the selecting parameter $a$ $(0 \le a \le 1)$
 to minimize Eve's mutual information of the shared key.

 When Eve performs only the storage attack,
 the best paramter is $a = 0$, that is, they use SARG protocol.
 On the other hand, when Eve performs the IRUD attack,
 the best is $a = 1$, that is, they use BB84 protocol.
\end{theorem}

\begin{proof}
 We will prove that the following equations:
 \begin{equation*}
 \begin{split}
  \partdif{a} I^{S}(a, \eta_{\rho}) &> 0 \\
  \partdif{a} I^{I}(a, \eta_{\rho}) &< 0 \text{.}
 \end{split}
 \end{equation*}

 We can calculate that
 \begin{equation*}
 \begin{split}
  \partdif{a} I^{S}(a, \eta_{\rho})
   &= ({\eta_{\rho}}^{-1} - 1) \cdot \partdif{a} \frac{I^{S}_{a}}{f(a)} \\
   &= ({\eta_{\rho}}^{-1} - 1)
       \cdot \frac{\partdif{a} I^{S}_{a} \cdot f(a) - I^{S}_{a} \cdot \partdif{a} f(a)}
                 {\{ f(a) \}^2}
 \end{split}
 \end{equation*}
 where $f(a) = \frac{2}{e^{- \mu_a} \mu_a} - 1$.

 Suppose that
 \begin{equation*}
 \begin{split}
  g(a)
   &= \partdif{a} I^{S}_{a} \cdot f(a) - I^{S}_{a} \cdot \partdif{a} f(a) \\
   &= (1 - L^{S}) \left( \frac{2}{e^{- \mu_a} \mu_a} - 1 \right) \\
   & \qquad \qquad  - (a + (1 - a) L^{S}) \frac{2 {\mu_a}' (1 - \mu_a)} {e^{- \mu_a} {\mu_a}^2}
 \end{split}
 \end{equation*}
 where $L^{S} = 1 - \mathcal{H} \left(\frac{\sqrt{2} + 1}{2 \sqrt{2}} \right)$
 and ${\mu_a}' = \partdif{a} \mu_a < 0$.

 Considering $L^{S}$ as variable, we can get
 \begin{equation*}
 \begin{split}
  \partdif{L^{S}} g(a)
   &= \frac{e^{- \mu_a} {\mu_a}^2 - 2 \mu_a + 2 (1 - a) {\mu_a}' (1 - \mu_a)}{e^{- \mu_a} {\mu_a}^2} \\
   &< \frac{2 \{ {\mu_a}^2 - \mu_a + (1 - a) {\mu_a}' (1 - \mu_a) \}}{e^{- \mu_a} \mu_a^2} \\
   &< \frac{2 \{ {\mu_a}^2 - \mu_a + (1 - \mu_a) \}}{e^{- \mu_a} \mu_a^2} \\
   &< \frac{2 \mu_a (\mu_a - 1)}{e^{- \mu_a} \mu_a^2} \\
   &\le 0
 \end{split}
 \end{equation*}

\begin{figure}
 \includegraphics[width=8.5cm, keepaspectratio, clip]{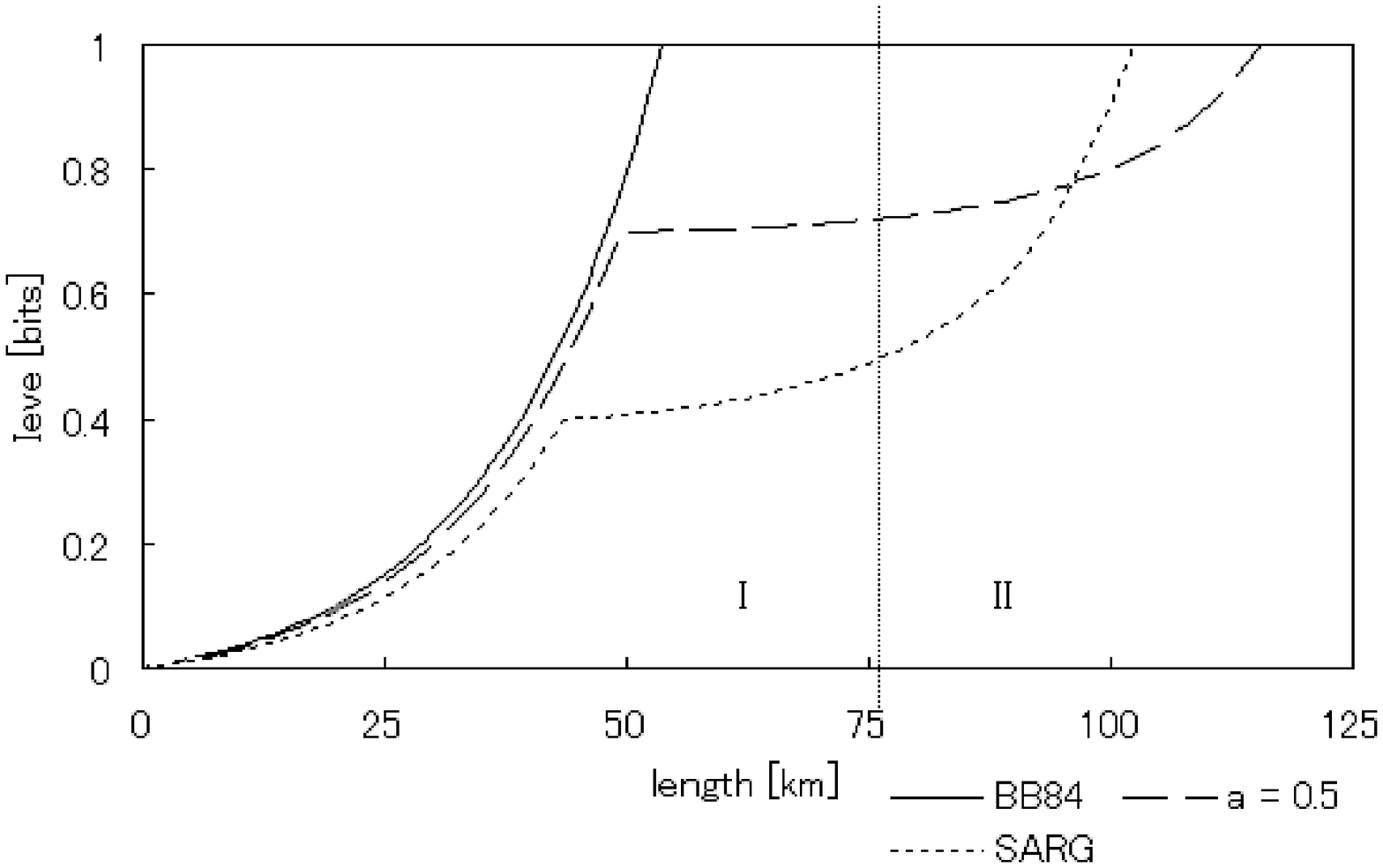}
 \caption{
  Security against PNS attacks with $\mathrm{QBER} = 0$.
  In the area I, Eve performs the storage attack and obtains an information about the sifted key.
  In the area II, Eve's attacks is shifted to the IRUD attack.
 }
 \label{fig:pns}
 \includegraphics[width=8.5cm, keepaspectratio, clip]{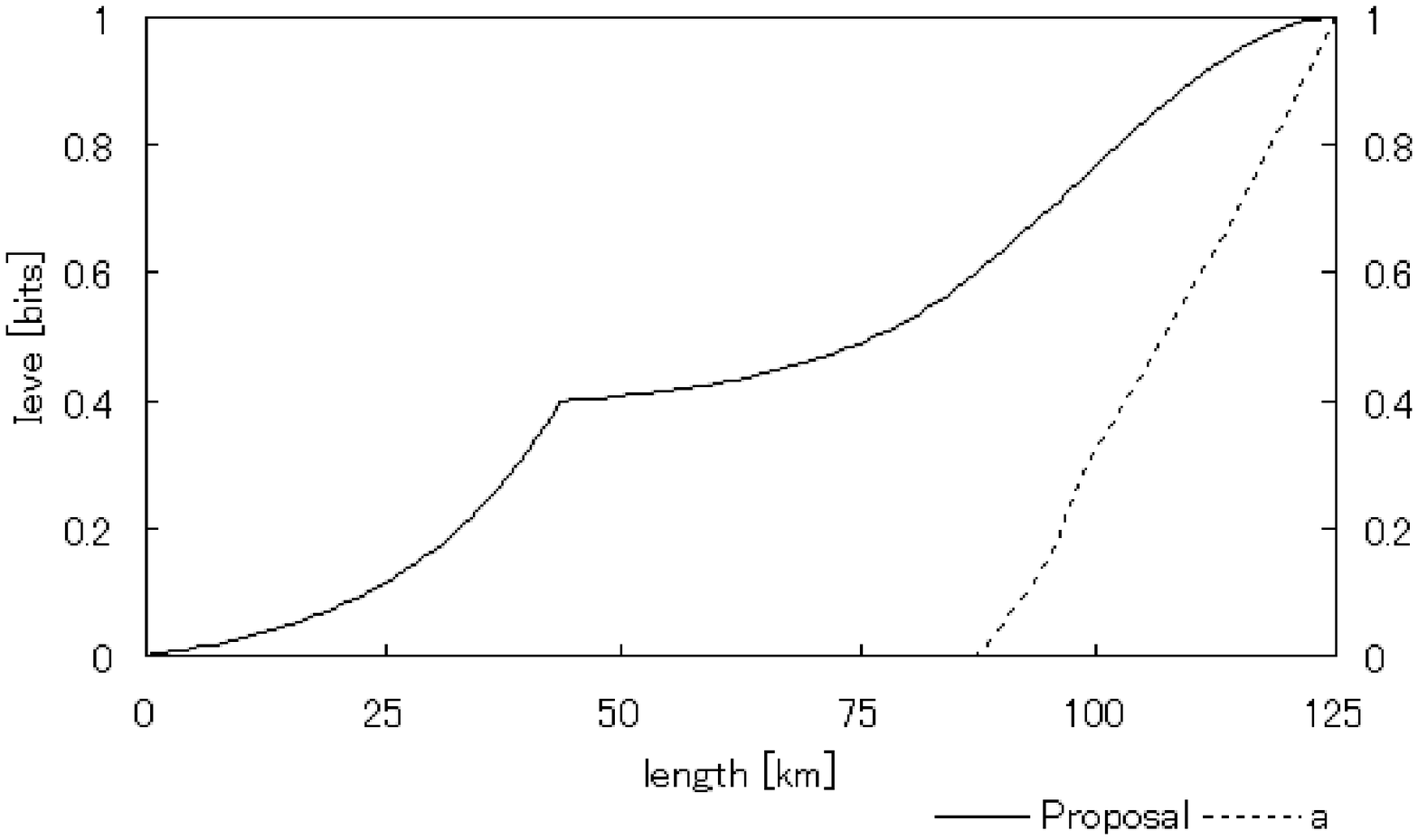}
 \caption{
  Security against PNS attacks when Alice and Bob choose an optimal $a$.
  If $l \le 87.5 \mathrm{km}$, they only use $a = 0$.
  If $l > 87.5 \mathrm{km}$, they increase $a$ shown in the dotted line.
  Comparing with Figure \ref{fig:pns},
  our protocol is more secure against PNS attacks than both of two protocols.
 }
 \label{fig:opt}
\end{figure}%

 Therefore, because $L^{S} < 0.5$, it can be shown that
 \begin{equation*}
 \begin{split}
  g(a)
   &> g(a)|_{L^{S} = 0.5} \\
   &= \frac{\mu_a (2 - e^{- \mu_a} \mu_a) - 2 (1 + a) {\mu_a}' (1 - \mu_a)}{2 e^{- \mu_a} {\mu_a}^2} \\
   &> \frac{\mu_a (1 - \mu_a) + (1 + a) {\mu_a}' (1 - \mu_a)}{e^{- \mu_a} {\mu_a}^2} \\
   &= 0
 \end{split}
 \end{equation*}
 where $(1 + a) {\mu_a}' = - (1 + a) \frac{2 \mu_B}{(1 + a)^2} = - \mu_a$.

 By ${\eta_{\rho}}^{-1} - 1 \ge 0$, we have $\partdif{a} I^{S}(a, \eta_{\rho}) > 0$.

 Next, suppose that
 \begin{equation*}
  \partdif{a} I^{I}(a, \eta_{\rho})
   = ({\eta_{\rho}}^{-1} - 1) \cdot - \frac{\partdif{a} h(a)}{\{ h(a) \}^2} \\
 \end{equation*}
 where $h(a) = \frac{12}{e^{- \mu_a} {\mu_a}^2} - 1$.

 Then
 \begin{equation*}
  \partdif{a} h(a) = \frac{12 {\mu_a}' (\mu_a - 2)}{e^{- \mu_a} \mu_a^3} > 0
 \end{equation*}
 because ${\mu_a}' < 0$ and $\mu_a < 2$.

 Therefore, $\partdif{a} I^{I}(a, \eta_{\rho}) < 0$.
\end{proof}

At Figure \ref{fig:pns}, we show Eve's maximal mutual information of a sifted key
when she performs either of two PNS attacks, as a function of the communication distance.
We use typical values $\eta_{\rho} = 10^{- \rho / 10}$, $\rho = \alpha l \mathrm{[dB]}$
and $\alpha = 0.25 \mathrm{[dB/km]}$, where $l$ is the length of the fiber.
We say that, in the case of $l \ge 100 \mathrm{km}$,
the proposed protocol with $a = 0.5$ is better than SARG protocol
because $I^{I}(0.5, \eta_{\rho}) \le I^{I}(1, \eta_{\rho})$.

Second, consider that Alice and Bob choose $a$ to minimize
Eve's mutual information when she performs the most convenient PNS attack,
in which her mutual information is
\begin{equation*}
 I^{P}(a, \eta_{\rho}) = \max \{ I^{S}(a, \eta_{\rho}), I^{I}(a, \eta_{\rho}) \} \text{.}
\end{equation*}
By Figure \ref{fig:opt}, we can say that, by choosing an optimal $a$,
the ultimate limit of robustness is shifted from $100 \mathrm{km}$,
which is the ultimate limit of SARG protocol, to $125 \mathrm{km}$,
which is the longest record among experimental QKD systems in the world.

\begin{acknowledgments}
This work was supported by the project on ``Research and Development on Quantum Cryptography''
of National Institute of Information and Communications Technology as part of the programme
``Research and Development on Quantum Communication Technology''
of the Ministry of Public Management, Home Affairs, Posts and Telecommunications Japan.
\end{acknowledgments}


\begin{thebibliography}{9}
\bibitem{AGS04}
 A. Ac\'{i}n, N. Gisin, V. Scarani, Phys. Rev. A \textbf{69}, 1 (2004)
\bibitem{BB84}
 C. H. Bennett, G. Brassard,
 in  \textit{Proceedings of the IEEE Conference on Computers, Systems and Signal Processing},
 \textit{Bangalore}, \textit{India} (IEEE, NewYork, 1984), pp. 175-179.
\bibitem{Lu00}
 N. L\"{u}tkenhaus, Phys. Rev. A \textbf{61}, 052304 (2000)
\bibitem{Per98}
 A. Peres, \textit{Quantum Theory: Concepts and Methods}, (Kluwer, Dordrecht, 1998), Sec. 9-5.
\bibitem{SARG04}
 V. Scarani, A. Ac\'{i}n, G. Ribordy, N. Gisin, Phys. Rev. Lett. \textbf{92}, 5 (2004)
\end{thebibliography}
\end{document}